\documentclass[twocolumn,showpacs,nofootinbib,10pt]{revtex4-1}

\usepackage{color}
\usepackage{graphicx}
\usepackage{graphicx,float}\usepackage[all]{xy}
\usepackage{amsmath,upgreek}
\usepackage{amssymb}

\newcommand{\beq}{\begin{eqnarray}}
\newcommand{\eeq}{\end{eqnarray}}
\newcommand{\be}{\begin{equation}}
\newcommand{\ee}{\end{equation}}
\newcommand{\bea}{\begin{eqnarray}}
\newcommand{\eea}{\end{eqnarray}}

\newcommand{\ba}{\begin{eqnarray}}
\newcommand{\ea}{\end{eqnarray}}
\usepackage[colorlinks,hyperindex]{hyperref}

\begin{document}
\title{Dark SU($N$) glueball stars on fluid branes}

\author{Rold\~ao da Rocha}
\affiliation{Centro de Matem\'atica, Computa\c c\~ao e Cogni\c c\~ao, Universidade Federal do ABC - UFABC\\ 09210-580, Santo Andr\'e, Brazil.}
\email{roldao.rocha@ufabc.edu.br}



\begin{abstract} 
The glueball dark matter, in the pure SU($N$)  Yang-Mills theory, engenders dark SU($N$) stars that comprise self-gravitating  compact configurations of scalar glueball fields.  
Corrections to the highest frequency of gravitational wave radiation  emitted by dark SU($N$) star mergers on a  fluid brane with variable tension, implemented by the minimal geometric deformation, are derived and their consequences analysed. Hence, dark SU($N$) stars mergers on a fluid brane-world are shown to be better detectable  by the LIGO and the eLISA experiments. 

\end{abstract}



\maketitle

\section{Introduction} 

Dark matter and dark energy comprise 
new directions  in gravity and high energy physics, towards theories that are beyond the General Relativity (GR) and which can explain such two puzzling  phenomena.
Among successful attempts to propose theories beyond GR, the method of geometrical deformation (MGD) consists of a suitable approach to derive new solutions of the effective Einstein field equations \cite{ovalle2007,covalle2,darkstars}, encoding compact stellar distributions. These new solutions are  complementary to other successful paradigms \cite{maartens,Antoniadis:1998ig}.  The MGD comprises the brane tension ($\sigma$) as a free parameter, controlling the high energy regime of an inflationary brane-world scenario that has the GR as the low energy limit. In fact, during the cosmological evolution, the brane temperature has been severely modified. It varied from, for instance,  $T \sim 10^4$ K -- when the matter density equaled the radiation density, around 5.6 $\times 10^3$ years after the Big Bang -- to the current value of $T\sim2.73$ K, in the CMB. An underlying 
setup can be, thus, implemented by a variable tension fluid brane \cite{Gergely:2008jr}, whereon  compact self-gravitating systems can undergo the MGD  \cite{Casadio:2013uma}.

The MGD was developed using Randall-Sundrum like models \cite{r1,maartens}. It has the bulk dark pressure and radiation as leading  constituents of the stress-energy tensor, in the brane effective Einstein field equations \cite{GCGR}. Explicit solutions are compact stellar structures with a bulk Weyl fluid imprint \cite{covalle2,darkstars,Ovalle:2007bn,ovalle2007}. 
 The MGD is a strong and robust procedure 
that has been recently endowed with observational and experimental precise bounds, from gravitational lensing effects \cite{Cavalcanti:2016mbe} to the classical tests of GR \cite{Casadio:2015jva}. 

On the other hand, gauge fields that are further away from the Standard Model of elementary particles were proposed as  pure Yang-Mills dark fields \cite{Juknevich:2009ji,Forestell:2016qhc}. In fact, the Standard Model can be coupled to  hidden sectors, governed by a pure Yang-Mills setup,  in the low energy regime. The scalar glueball dark matter model implements such a SU($N$) Yang-Mills sector \cite{Juknevich:2009ji,Yamanaka:2014pva,Soni:2016gzf,Boddy:2014yra}, as a  self-interacting field with large
cross section. 
When Standard Model  particles and fields are forbidden to interact with the SU($N$) scalar glueball, gravity  can induce a self-gravitating system, manifesting Bose-Einstein condensation of glueballs 
  into compact stellar objects.  
  Ref. \cite{Soni:2016gzf} discusses relevant elastic scatterings among glueballs, manifesting their feasibility as a self-interacting dark matter candidate. A refined and detailed analysis can be checked in Ref. \cite{Juknevich:2009ji}, together with other general aspects \cite{Miranda:2009uw}.

Although dark SU($N$) compact systems were studied \cite{Soni:2016gzf}, any realistic approach that provides  observational and experimental signatures of dark SU($N$) stars on inflationary scenarios lacks, still. 
Here  the MGD is proposed as a procedure 
to implement  dark SU($N$) brane-world stars, in the context of the 
evolution of the Universe, as well as to refine the analysis of gravitational waves produced by  dark SU($N$) glueball star mergers, enhancing the window for current experiments to detect them.

This paper is organised as follows: Sect. II is devoted to a brief review, regarding the MGD of stellar distributions on a fluid brane, ruled by a variable tension that encodes the cosmological evolution. In Sect. III, corrections to the highest frequency of gravitational wave radiation,  emitted by dark SU($N$) star mergers  due to the finite brane tension, are derived for both the $\phi^4$ self-interacting glueball potential and the glueball potential in the large $N$ limit. 
These corrections make the gravitational wave radiation, emitted by dark SU($N$) MGD stars mergers, to be more able to detect than in the GR limit setup. Hence, a larger spectrum of gravitational waves on fluid branes is expected, enhancing the window to be probed by the LIGO and the eLISA experiments. Sect. IV is dedicated to draw the conclusions and final comments.

\section{The MGD setup and fluid branes}
\label{MGD}
 The MGD procedure is able to derive high energy corrections to GR, when the vacuum in the outer region of a compact distribution is permeated  by a 5-dimensional (5D) bulk Weyl fluid \cite{ovalle2007,Ovalle:2013xla,darkstars}. 
The codimension-1 brane that designates our Universe has tension (self-gravity) as a leading  parameter, which varies as the temperature decreases, across the Universe  inflation  \cite{Gergely:2008jr,Abdalla:2009pg}. The most useful and applicable brane-world 
scenarios in this context are implemented by  fluid branes, evincing the E\"otv\"os law that provides the  dependence of the brane tension with the temperature  \cite{Gergely:2008jr,Abdalla:2009pg}.

The MGD setup has recently imposed the  brane (variable) tension bounds  $\sigma \gtrsim  5.19\times10^6 \;{\rm MeV^4}$  (in the context of the classical tests of GR) \cite{Casadio:2015jva} and $\sigma \gtrsim  3.18\times10^6 \;{\rm MeV^4}$ (regarding the Bose-Einstein condensation of weakly interacting gravitons into MGD black holes)  \cite{Casadio:2016aum}. The MGD represents a deformation of the Schwarzschild metric, implemented by bulk effects in the brane-world paradigm, whose low energy regime $\sigma\to \infty$  recovers the Schwarzschild standard solution.

The 4D Einstein effective equations can be derived when the 5D bulk Einstein equations are projected onto the 4D brane, by the Gauss-Codazzi method  (the convention  $8\pi G=c=1=\hbar$ is going to be fixed\footnote{Obviously, the precise values of all involved  parameters shall 
be suitably taken into account, in the calculations in Sect. III.}, where $G = \hbar c/M_{\rm pl}^2$ and $M_{\rm pl}$ denotes the Planck scale, and ${\footnotesize{\mu,\nu}}=0,1,2,3$), yielding ~\cite{GCGR} 
\begin{equation}
\label{5d4d}
R_{\mu\nu}+\left(\Lambda-\frac12 R\right)g_{\mu\nu}
-{\rm T}_{\mu\nu}=0,
\end{equation} where $R_{\mu\nu}$, $R$, and $\Lambda$ are, respectively, the Ricci tensor, the Ricci scalar, and the 4D cosmological constant. 
The effective stress tensor in Eq. (\ref{5d4d}) can be split into a sum, 
$
{\rm T}_{\mu\nu}
=
T_{\mu\nu}+\mathcal{E}_{\mu\nu}+\sigma^{-1}S_{\mu\nu}
$, 
where the first component $T_{\mu\nu}$ denotes the brane matter stress tensor and $\mathcal{E}_{\mu\nu}$ inscribes high-energetic corrections from the 5D Weyl fluid. The tensor  
$S_{\mu\nu}$ encrypts Kaluza-Klein imprints from the bulk onto the brane \cite{GCGR,maartens}.

Compact stellar structures that are solutions of the Einstein brane field equations  (\ref{5d4d}) are usually obtained for static, spherically symmetric, metrics (\ref{abr}), \begin{equation}\label{abr}
ds^{2} = -A(r) dt^{2} + (B(r))^{-1}dr^{2} + r^{2} (d\vartheta^2+\sin^2\vartheta d\varphi^2). 
\end{equation} 
The MGD procedure fixes  
the $g_{tt}$ metric component in (\ref{abr}) 
and deforms  the outer $g_{rr}^{-1}\equiv B(r)$ metric component  \cite{covalle2,darkstars,Ovalle:2007bn},  
\begin{eqnarray}
\label{deff}
B(r)
=
{1-\frac{2\,M}{r}}
+\varsigma\,\exp\left(
\int^r_{{\rm R}}
\frac{f(A({\rm r}))}{g(A({\rm r}))}\,d {\rm r}\right)
\ ,
\label{I}
\end{eqnarray}
where $f(A(r))\equiv \frac{AA''}{A'^{2}}\!+\!(\ln(A))^{\prime2}+\!\frac{2}{r}\ln(A)^{\prime}\!-1+\!\frac{1}{r^{2}}$ and 
$g^{-1}(A(r))=\frac12\ln(A)^{\prime}+\frac{2}{r}$ \cite{covalle2},  ${(\;\;\;)}^\prime\equiv\frac{d(\;\;\;)}{dr}$, and  ${\rm R} \equiv {\int r^3\,\rho(r)\,  dr}/{\int r^2 \rho(r)\,dr},$ where $\rho$ is the
energy density of the stellar matter distribution \cite{ovalle2007}. The parameter $\varsigma$ in Eq. (\ref{deff}) encodes a  bulk-induced deformation of the vacuum,
 at the compact distribution surface, comprising the necessary 5D (bulk) Weyl fluid data onto the brane~\cite{Casadio:2015jva}. 
It is significant to observe that the outer metric is defined in the region $r>{\rm R}$~\cite{ovalle2007}, whose  deformation yields ~\cite{covalle2}
\begin{subequations}
\ba
\label{nu}
\!\!\!\!\!\!A(r)
&=&
1-\frac{2\,M}{r}
\ ,
\\
\!\!\!\!\!\!B(r)
&=&
\left[1+\left({1-\frac{3\,M}{2\,r}}\right)^{-1}\!\!{\varsigma}\frac{\mathfrak{l}}{r}\,\right]\left(1-\frac{2\,M}{r}\right)
\ ,
\label{mu}
\ea
\end{subequations} 
where 
\begin{equation}
\label{L}
\mathfrak{l}
\equiv
\left(1-\frac{2M}{{\rm R}}\right)^{\!-1}\!
\left(1-\frac{3M}{2{\rm R}}\right){\rm R}.
\end{equation}
{The MGD black hole event horizons are $r_1 = 2\,M$ and  
$r_2=- \varsigma\,\mathfrak{l}+\frac{3M}{2}$.  
The infinitely rigid brane limit $\varsigma^{-1}\sim \sigma\to\infty$, that characterises the GR limit, yields   $r_1> r_2$. 
Besides, $\varsigma$ is a parameter that relies on the inherent compact star configuration. Refs. \cite{ovalle2007,Casadio:2013uma} show that the metric radial component (\ref{nu}) can be written as 
\begin{eqnarray}
\!\!\!\!\!\!\!B(r)=1-\frac{2M}{r}-\left({1-\frac{2M_0}{r}}\right)\left({1-\frac{3M_0}{2\,r}}\right)^{-1}\frac{\mathfrak{l}_0\,\varsigma}{r}\,,
\end{eqnarray} thus exhibiting a part that is beyond the Schwarzschild solution,  
up to order ${\cal O}(\sigma^{-2})$, where $M_0 = M\vert_{\sigma\to\infty}$ is the GR mass function and $\mathfrak{l}_0=\mathfrak{l}(M_0)$, concerning  Eq.~(\ref{L}).
Bulk imprints are highest at the star surface
$r={\rm R}$. This shall be a prominent information in the analysis of dark SU($N$) stars on fluid branes, in the next section. 

The less compact the star, the smaller the $|\varsigma|$ parameter is 
\cite{ovalle2007,covalle2}. 
Moreover, the current experimental and observational data was used in Ref. \cite{Casadio:2015jva}, to impose the strongest bound $
|\varsigma|\lesssim 6.1 \times 10^{-11}$ on the MGD parameter, using the classical tests of GR.   
This result, together with the most recent and strict bound on the variable brane tension $\sigma \gtrsim  3.18\times10^6 \;{\rm MeV^4}$ \cite{Casadio:2016aum}, justifies high order ${\cal O}(\sigma^{-2})$ terms to be dismissed, yielding
\begin{eqnarray}
\label{cobsigma}
\varsigma(\sigma,{\rm R})
= 
-\frac{b_0}{{\rm R}^2}\sigma^{-1}
\ ,
\end{eqnarray} where $b_0\sim  1.35$ \cite{Casadio:2015jva}. 
The negativeness of $\varsigma$ compels the  MGD star gravitational field to be weakened, as an effect of a finite brane tension, when compared to the GR regime $\sigma\to\infty$.
}

Heretofore no condition on the 
variable brane tension has been imposed, although cosmological evidence drives the brane tension to fluctuate across  the Universe inflation \cite{Gergely:2008jr,Wong:2010rg}. On E\"otv\"os fluid branes, the brane tension varies with respect to the Universe temperature $T$. One associates the regime  $\sigma\approx T-\tau$ \cite{Gergely:2008jr}, where  $\tau$ is some critical value that makes $\sigma$ to assume only non-negative values after the Big Bang \cite{Gergely:2008jr,Abdalla:2009pg}. This varying brane tension can eliminate any initial singularity at the early Universe. In fact, the brane Universe was created at a $\tau$ temperature, corresponding to the scale factor value $a_{0}$ that is defined by the coupling constants  \cite{Gergely:2008jr,Wong:2010rg}. Ref. \cite{Gergely:2008jr} derived the relationship between temperature and the scale factor, $T(t) \approx \frac{1}{a(t)}$ \cite{Gergely:2008jr}. This result yields a time-dependent brane tension \begin{eqnarray}\label{tensao}\frac{\sigma(t)}{\sigma_{0}}=1-\frac{a_{0}}{a(t)}.\end{eqnarray}  At the extremely hot early Universe, the brane tension was negligible ($\sigma\approxeq0$). Subsequently,  both the variable brane tension and the 4D coupling parameter grew, as the scale factor asymptotically increased, in the inflationary brane-world scenario \cite{Halverson:2016nfq}. 
The time-dependent brane tension expression yields $\frac{\Lambda_{{\rm 4D}}}{\Lambda_0}=1-\frac{a_0}{a(t)}\left(1-\frac{a_0}{a(t)}\right)$. 
This (dynamical) cosmological ``constant'' had attained a huge negative value and  achieved lowly positive values \cite{Abdalla:2009pg}. It also engenders supplementary attraction [repulsion] at small [large] values of the scale factor, similarly to the dark matter [dark energy]. This inflationary cosmology scenario emulates the 
  (cosmological) standard model at late times, wherein the  energy that is absorbed by the  brane thrusts the 5D bulk towards a anti-de Sitter bulk, as a mere consequence of highest symmetry \cite{Wong:2010rg,Barosi:2008cs}.

In the next section, the dark SU($N$) stellar system, constituted of self-interacting scalar glueballs, is studied on a fluid brane, whose   tension obeys the E\"otv\"os  law. Corrections to the gravitational waves frequency, emitted by   
dark SU($N$) stars mergers on the fluid brane, are then derived as a consequence of the dark SU($N$) stars mass and radii variation on a fluid brane. Moreover, dark SU($N$) stars features can be, in this context, analyzed along the  inflationary brane-world era.

\section{Fluid brane corrections to dark SU($N$) stars}

Hidden  SU($N$) Yang-Mills sectors, that are further away the Standard Model of elementary particles, can be realized by the (scalar) glueball dark matter model \cite{Juknevich:2009ji,Soni:2016gzf}. The gravitational interactions among glueballs engender a self-interacting system, having the glueball mass ($m$) and the number ($N$) of colors as the main underlying parameters.
When $10$ eV $\lesssim  m \lesssim$ 10 KeV  and $10^3 \lesssim N \lesssim 10^6$, then  the dark glueball field self-gravity takes part, forming dark SU($N$) compact stellar distributions \cite{Soni:2016gzf,Forestell:2016qhc}. 
In order to be compatible with 
the Bose-Einstein condensation of the scalar glueballs into static, spherically, symmetric solutions, 
the glueball fields are assumed to be periodic, $\phi(r,t) = \Phi(r)\cos\omega t$  \cite{Soni:2016gzf}. 

The action for the most general nonlinear interacting scalar field theory reads 
\begin{equation}
\mathcal{S} =  \int d^4x\,\left(\frac{1}{2} g^{\mu\nu} \partial_\mu \phi \partial_\nu\phi - V(\phi)\right).
\end{equation}
The Klein-Gordon equation, that is derived from this action by the Euler-Lagrange equations, couples to the Einstein field equations and shall be analyzed, for both the large $N$ glueball potential and for the $\phi^4$ self-interacting glueball potential as well. Hereon $\dot{(\;\;\;)}\equiv\frac{d(\;\;\;)}{dt}$. 

In the next two subsections,  corrections due to the variable brane tension, to the highest frequency of the gravitational wave radiation emitted by SU($N$) MGD dark stars mergers, shall be analyzed.  
Such a highest frequency is given, for the Schwarzschild case, by \cite{Soni:2016gzf} $f_{\rm max} =\frac{1}{2\pi}\left(\frac{GM}{R^3}\right)^{1/2}$, that can be probed by the LIGO experiment \cite{Abbott:2007kv}. SU($N$) MGD dark stars mergers are, then, shown to provide a better detection of gravitational waves, that is compatible with 
the current bounds of the brane tension and the CMB, in an inflationary brane-world scenario. 
\vspace*{-.6cm}
\subsection{Self-interacting $\phi^4$  glueball potential}\vspace*{-.2cm}
A   self-interacting potential for the dark glueball, 
\begin{equation}\label{quartic}
V(\phi) = \frac{1}{2} m^2 \phi^2 + \frac{\lambda}{4!}  \phi^4 \ ,
\end{equation}  was shown to rule both the glueball dark matter self-interaction strength and the properties of  dark SU($N$) stars as well. 
For $\lambda>0$, stable dark SU($N$) star configurations can be derived, when the repulsive $\lambda \phi^4$ self-interaction compensates gravity \cite{Soni:2016yes,Soni:2016gzf}. 

The coupled system of Einstein equations and Klein-Gordon  ones was derived in \cite{Soni:2016gzf}. Imposing time averaging over the  oscillation period $\frac{2\pi}{\omega}$ of the  $\phi$ field, such system yields  
\begin{widetext}
\begin{subequations}
\begin{eqnarray}
M'(x) - \left[\frac{\Lambda}{6}  \Phi^4(x)+ \left(1+\frac{\Omega^2}{A(x)} \right) \Phi^2(x)  + \frac{\Phi^{\prime2}(x)}{B(x)} \right]\frac{x^2}{8} &=&0 \ , \\
\frac{\left(\ln A(x)\right)'}{B(x)}  + \frac{\Lambda}{48}  x\Phi^4(x)+ \frac{x}{4} \left(1- \frac{\Omega^2}{A(x)} \right) \Phi^2(x)  - \frac{x}{8}\frac{\Phi^{\prime2}(x)}{B(x)}+ \frac{1}{x}\left(\frac{1}{B(x)}-1 \right)&=&0 \ , \\
\Phi''(x) + \left( \frac12\left(\ln\frac{A}{B}\right)' +\frac{2}{x}\right) \Phi'(x) -  \left[ \frac{\Lambda}{2}  \Phi^2(x)+ \left(1-\frac{\Omega^2}{A(x)} \right) \right]B(x)\Phi(x)  &=&0 \label{kg2} \ ,
\end{eqnarray}
\end{subequations}
\end{widetext}
where \cite{Soni:2016yes,Soni:2016gzf} 
\begin{subequations}
\begin{eqnarray}
x&=& m r,\\
\Lambda &=& \frac{12\lambda}{m^2},\\\Omega &=& \frac{\omega}{m}.
\end{eqnarray}
\end{subequations}
The regime $\Lambda \gg 1$ holds  for the glueball dark matter model  \cite{Colpi:1986ye}, yielding Eq. (\ref{kg2}) to imply 
\begin{eqnarray}
&&\Lambda^{-1}\left[\upphi''({\rm x}) + \left( \frac{2}{{\rm x}} + \frac12 \ln\left(\frac{A}{B}\right)^\prime \right) {\upphi}'({\rm x})\right] \nonumber\\&&- B({\rm x})\upphi({\rm x}) \left[ \left(1-\frac{\Omega^2}{A({\rm x})}  \right) - \frac{1}{2} \upphi^2({\rm x}) \right] =0 \ , \label{Klein1} 
\end{eqnarray} where 
\begin{eqnarray}
{\rm x} = \frac{x}{\sqrt{\Lambda}},\qquad \upphi = \sqrt{2\Lambda} \Phi,\qquad
{\rm M} = \frac{M}{\sqrt{\Lambda}}.
\end{eqnarray}
The $\Lambda\gg1$ limit can induce the  first term in Eq. (\ref{Klein1}) to be dismissed, yielding   
\begin{eqnarray}\label{limite}
\lim_{\Lambda\gg1}\left[\upphi({\rm x}) - \sqrt{ 2} \left(\frac{\Omega^2}{A({\rm x})} - 1 \right)^{1/2}\right]=0 \label{Klein2} \,,
\end{eqnarray}implying  that 
\begin{eqnarray}
\!\!\!\!&&\!\!\!\!{\rm M}'({\rm x}) \!-\! {\rm x}^2 \left[ \frac{1}{4} \left( \frac{\Omega^2}{B({\rm x})} \!+\! 1 \right) \upphi^2({\rm x}) \!+\! \frac{3}{32} \upphi^4({\rm x}) \right]=0 \ , \label{einstein1} \\
\!\!\!\!&&\!\!\!\!B(\ln A)^\prime{\rm x}^2 -2 {\rm M}+
\left[\left(1\!-\! \frac{\Omega^2}{A}\right)\frac{\upphi^2}{2} \!+\! \frac{3}{16}\! \upphi^4\right]{\rm x}^3=0, \label{einstein2}
\end{eqnarray} where ${\rm x}$ is the argument of all functions in Eqs. (\ref{einstein1}, \ref{einstein2}). Similarly to Refs. \cite{Soni:2016yes,Soni:2016gzf}, these equations  can be  solved by numerical methods, with ${\rm M}(0)=0$, for $0\leq {\rm x}\leq x_R$. 

The results obtained from using the self-interacting $\phi^4$ glueball potential  (\ref{quartic})
 are depicted in Figs.~1 -- 3.  In what follows, $\sigma \sim 10^6 \;{\rm MeV^4}$ shall denote the 
current brane tension bound $\sigma \gtrsim  3.18\times10^6 \;{\rm MeV^4}$ \cite{Casadio:2016aum}. The same notation shall be used for the values $\sigma \sim10^9$ and $10^{12}\sim{\rm MeV^4}$, but these exact values shall be adopted in what follows, unless otherwise explicitly stated.
 \begin{figure}[H]\label{1144}
\centering\includegraphics[width=7.1cm]{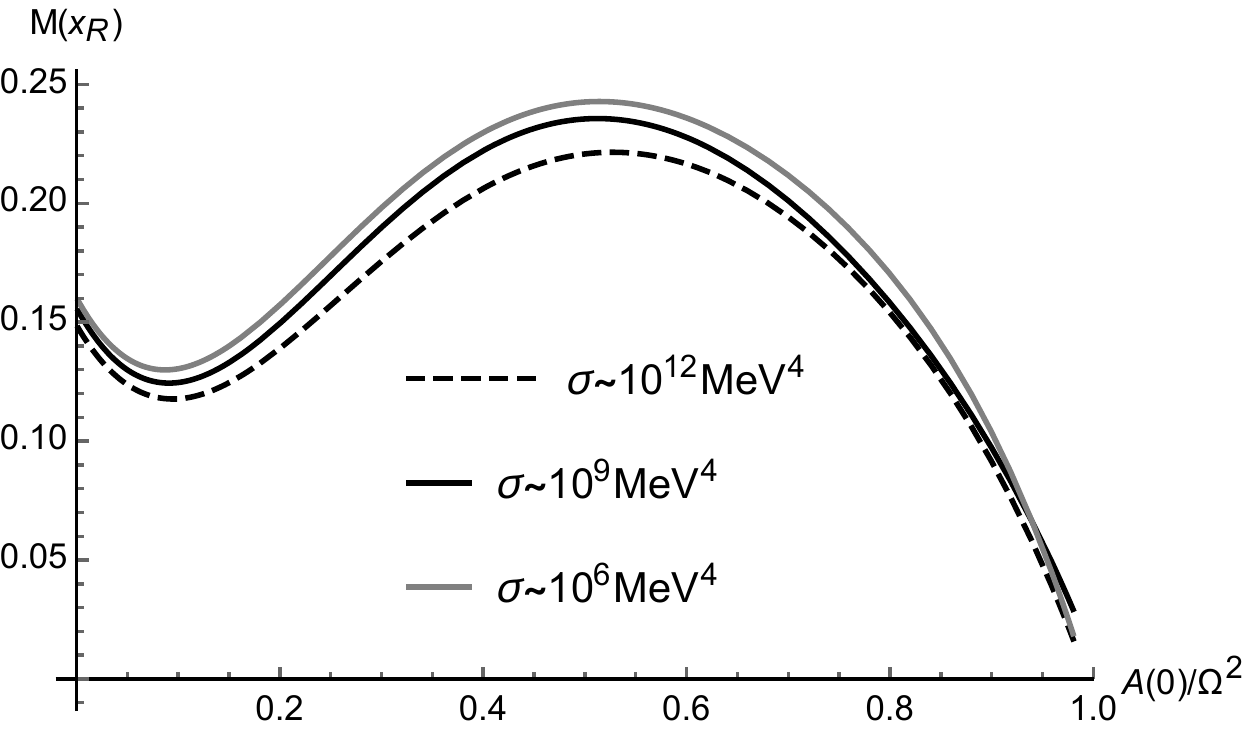}
\caption{Dark SU($N$) MGD star mass ${\rm M}(x_R)$, in the $\phi^4$ scalar glueball potential setup, normalized by $\frac{\sqrt{2\lambda}\,M_{\rm pl}^3}{m^2}$, with respect to $\frac{A(0)}{\Omega^2}$, for different values of the fluid brane tension $\sigma = 10^{12}$ MeV$^4$ (dashed line); $\sigma = 10^9$ MeV$^4$ (black line); $\sigma \sim 10^{6}$ MeV$^4$ (gray line).}
\end{figure}\begin{figure}[H]
\centering\includegraphics[width=7.1cm]{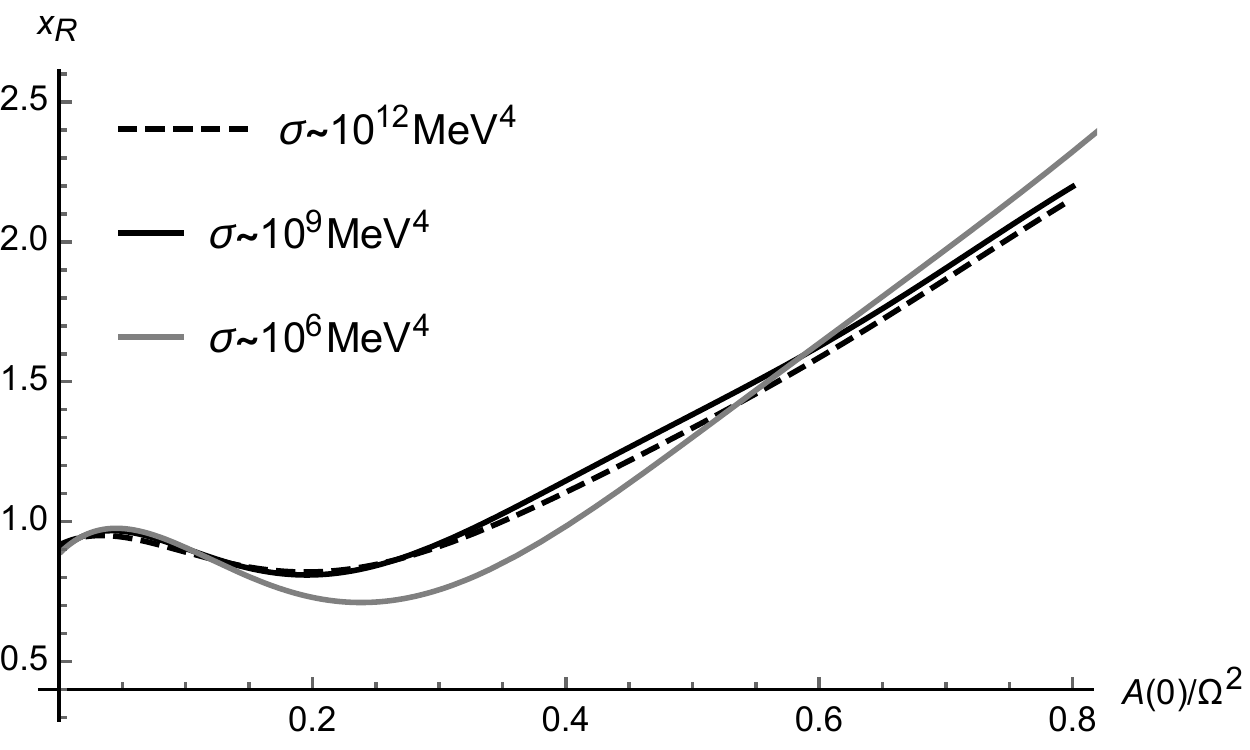}
\caption{Dark SU($N$) MGD star radius $x_R$ in the $\phi^4$ scalar glueball potential setup, normalized by $\frac{\sqrt{2\lambda}\,M_{\rm pl}}{m^2}$, with respect to $\frac{A(0)}{\Omega^2}$, for different values of the fluid brane tension $\sigma = 10^{12}$ MeV$^4$ (dashed line); $\sigma = 10^9$ MeV$^4$ (black line); $\sigma = 10^{6}$ MeV$^4$ (gray line).}
\end{figure}
 \begin{figure}[H]
\centering\includegraphics[width=7.1cm]{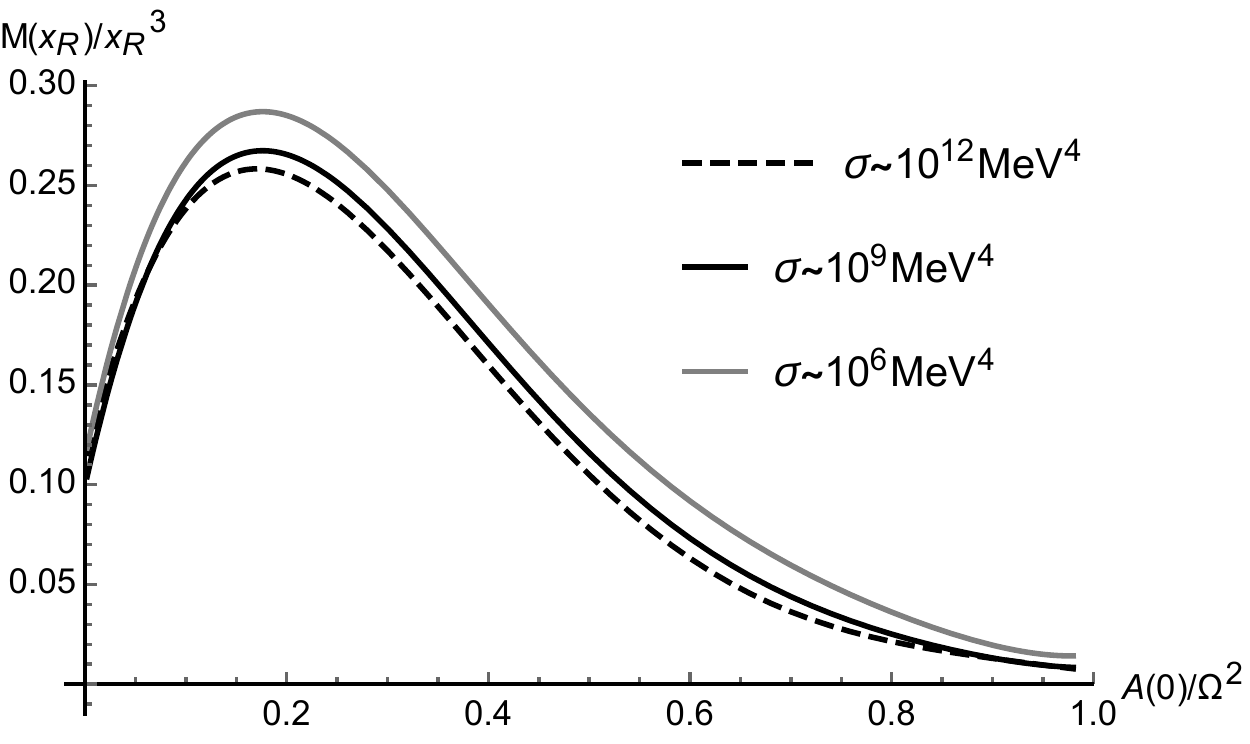}\caption{Dark SU($N$) MGD star ratio $\frac{{\rm M}(x_R)}{x_R^3}$ in the $\phi^4$ scalar glueball potential setup, normalized by $\frac{\sqrt{2\lambda}\,M_{\rm pl}}{m^2}$, with respect to  $\frac{A(0)}{\Omega^2}$, for different values of the fluid brane tension $\sigma = 10^{12}$ MeV$^4$ (dashed line); $\sigma = 10^9$ MeV$^4$ (black line); $\sigma \sim 10^{6}$ MeV$^4$ (gray line).}\label{133}
\end{figure}

The gravitational balance of self-interacting scalar fields was studied \cite{Colpi:1986ye}, also in the context 
of stability bounds on compact objects \cite{Gleiser:2013mga,Casadio:2016aum}, represented by spherically symmetric boson star solutions. 
According to the $g_{rr}^{-1}$ metric component in Eq. (\ref{mu}), 
the boundary condition ${\rm M}({\rm x}=0) = 0$ implies that $\lim_{{\rm M}\to 0}B(r)=1+\varsigma\frac{\mathfrak{l}}{r}$.

In Fig. 1,  the mass spectrum ($y$-axis) between  the point $\frac{A(0)}{\Omega^2}=0$ up to the critical (maximum) point in the plots cannot be attained. In fact, Ref. \cite{Soni:2016yes} 
showed that a dark SU($N$) star accretes by seizing the surrounding dark matter.
Thereafter, the dark SU($N$) MGD star mass increases by accretion, up to a maximum, represented in the third column in Table I. The results are presented for different values of the brane tension: 
\begin{center}
\begin{table}[!h]
\begin{tabular}{||c||c||c||c||c||}\hline\hline
\;Brane tension $\sigma$\vspace{0.06cm}\;&\;$
\frac{A(0)}{\Omega^2}$\;& Mass M$(x_R)$& Radius $\;x_R$&$\frac{{\rm M}(x_R)}{x_R^3}$\\\hline\hline
$\infty$ (GR limit)&0.533&0.222&1.351&0.090\\\hline
$10^{12}$  (MeV$^4$)&0.530&0.223&1.392&0.091\\\hline
$10^{9}$  (MeV$^4$)&0.517&0.240&1.421&0.110\\\hline
$10^{6}$  (MeV$^4$)&0.509&0.249&1.381&0.142\\\hline
\hline
\end{tabular}
\caption{highest values of the dark SU($N$) MGD stars radius $\left(\text{normalized by $\frac{\sqrt{2\lambda}}{m^2}\,M_{\rm pl}$}\right)$ and mass $\left(\text{normalized by $\frac{\sqrt{2\lambda}}{m^2}\,M_{\rm pl}^3$}\right)$,  by accretion, for different values of the fluid brane tension, for a $\lambda\phi^4$ scalar glueball potential.}
\end{table}
\end{center}
\newpage
Now, the glueball SU($N$) dark star has radius and mass that read, respectively  \cite{Soni:2016yes,Soni:2016gzf}, 
\begin{eqnarray}\label{massa}
R &=&   \frac{\sqrt{2\lambda}}{m^2} M_{\rm pl}\,x_R,\\
M &=&  \frac{\sqrt{2\lambda}}{m^2}\,M_{\rm pl}^3\, {\rm M}(x_R).\label{raioo}
\end{eqnarray}
Based on the third and fourth columns in Table I, the glueball dark SU($N$) MGD star has highest radius and mass, respectively, given by:
\begin{eqnarray}\label{massa1}
\!\!\!\!\!\!\!\!\!\!\!\!\!\!\!\!\!\!\!\!\!\!\!\!\!\!\!\!\!\!\!\!\!\!\!\!\!R\!= \begin{cases}902.5 \;m^2 \sqrt{\lambda},&\!\!\quad \text{for $\sigma\to\infty$ (GR limit)}\label{mm1}\\
929.8 \;m^2 \sqrt{\lambda},&\!\!\quad \text{for $\sigma=10^{12}\;{\rm MeV}^4$} \label{mfka2}\\
949.2 \;m^2 \sqrt{\lambda},&\!\!\quad \text{for $\sigma=10^{9}\;{\rm MeV}^4$}\label{mfka3}\\
922.5 \;m^2 \sqrt{\lambda},&\!\!\quad \text{for $\sigma\sim10^{6}\;{\rm MeV}^4$}\label{mfka4}
\end{cases}\\
\!\!\!\!\!\!\!\!\!\!\!\!M\!= \begin{cases}\frac{9\sqrt{\lambda}}{m^2}10^{-2}M_{\odot},&\!\!\quad \text{for $\sigma\to\infty$ (GR limit)}\label{mm1}\\
\frac{9.04\sqrt{\lambda}}{m^2}\;10^{-2}  M_{\odot},&\!\!\quad \text{for $\sigma=10^{12}\;{\rm MeV}^4$} \label{mfka2}\\
\frac{9.72\sqrt{\lambda}}{m^2}\;10^{-2}  M_{\odot},&\!\!\quad \text{for $\sigma=10^{9}\;{\rm MeV}^4$}\label{mfka3}\\
\frac{10.94\sqrt{\lambda}}{m^2}\;10^{-2}  M_{\odot},&\!\!\quad \text{for $\sigma\sim10^{6}\;{\rm MeV}^4$}\label{mfka4}
\end{cases}\label{raioo1}\end{eqnarray} 
where $M_{\odot}$ denotes, as usual,  the Solar mass.

 Eqs.~(\ref{massa1}, \ref{raioo1}), together with the last column of Table I, yield the highest gravitational wave radiation frequency,
\begin{eqnarray}\label{aaaa}
\!\!\!\!\!\!\!\!\!\!\!\!\!\!\!f_{\rm max} = \frac{m^2\sqrt{\lambda}}{\sqrt{2}\pi M_{\rm pl}}  {\rm supp}^{1/2}\!\left(\frac{{\rm M}(x_R)}{x_R^3}\right)\approxeq
\beta_1(\sigma) (50\, {\rm Hz}),
\end{eqnarray} 
where the function 
\begin{eqnarray}
 \beta_1(\sigma)=  \mathfrak{a}\sqrt\lambda (2m)^2\;10^4\; {\rm GeV}^{-2},
\end{eqnarray}
 has an adjusting factor $\mathfrak{a}$, that is a function of the variable brane tension, according to Eq. (\ref{aaaa}) when the last column of Table I is taken into account. Such factor is given by
\begin{eqnarray}\label{mfka1}\mathfrak{a}=
\begin{cases}1,&\quad \text{for $\sigma\to\infty$ (GR limit)},\label{mfka1}\\
1.010,&\quad \text{for $\sigma=10^{12}\;{\rm MeV}^4$}, \label{mfka2}\\
1.111,&\quad \text{for $\sigma=10^{9}\;{\rm MeV}^4$},\label{mfka3}\\
1.262,&\quad \text{for $\sigma\sim10^{6}\;{\rm MeV}^4$}\label{mfka4}.
\end{cases}\end{eqnarray}
The parameter $\mathfrak{a}$ indicates 
the corrections to the unit (that corresponds to the $\sigma\to\infty$ GR limit), for different values of the brane tension. 

The parameter $\mathfrak{a}$ in (\ref{mfka1}) shows corrections to the highest gravitational wave radiation frequency emitted by  dark SU($N$) MGD star mergers, due to the brane tension. The current lower bound for the brane tension $\sigma \gtrsim  3.18\times10^6 \;{\rm MeV^4}$ \cite{Casadio:2016aum}, in the last line of Eq. (\ref{mfka1}), provides a  realistic fluid brane scenario, wherein the highest gravitational wave radiation frequency is up to $\sim 26.2\%$ higher than the predictions in the GR limit. 

Now, we can see that the highest gravitational wave frequency that is emitted by dark SU($N$) MGD star mergers, in Eqs.~(\ref{aaaa}), for the self-interacting $\phi^4$ glueball potential, can be better detected by the LIGO and the eLISA experiments \cite{Abbott:2007kv}, having a wider range than the spectrum of frequencies provided by Schwarzschild solutions \cite{Soni:2016gzf}. This shall be clear in what follows, by analyzing the $N$ - $m$ parameter space.

Ref. \cite{Soni:2016gzf} argued that the dark SU($N$) stars have parameters in the ranges $100$ eV $\lesssim m \lesssim 10\,$KeV and $10^3\lesssim N \lesssim 10^6$, yielding a maximum $10^6M_{\odot}\lesssim{M}\lesssim10^9M_{\odot}$ for the dark SU($N$) MGD star mass, whereas the lowest dark SU($N$) MGD star radius varies in the range $10^2\lesssim R\lesssim10^5$, in unit of the Solar radius $R_\odot$. 
Hence,  the highest gravitational wave frequency that is emitted by dark SU($N$) MGD stars mergers, given by Eqs.~(\ref{aaaa}), for the $\phi^4$ glueball potential (\ref{quartic}), can be   better detectable by the LIGO and the eLISA experiments \cite{Abbott:2007kv}. Moreover,  dark SU($N$) MGD star mergers have specific 
signatures that are quite different of Schwarzschild black hole mergers, due to the subsequent analysis of Table I, as well as Eqs. (\ref{mfka1}). In fact, 
since dark SU($N$) MGD stars do not necessarily collapse to form a black hole, their gravitational wave frequency of emission has a distinct signature of the ones emitted by black hole mergers.

The  highest frequency of gravitational wave radiation $f_{\rm max}$, emitted from dark SU($N$) MGD stellar mergers, can be allocated in the range 30 $\mu${\rm Hz} $\lesssim f_{\rm max}\lesssim 100$ mHz, that can be further detected by the eLISA mission \cite{Seoane:2013qna}. In addition, the LIGO experiment  can  probe the range of gravitational waves frequency 50 {\rm Hz} $\lesssim f_{\rm max}\lesssim 1$ KHz, nowadays. Both these  ranges are, respectively, represented by the light-gray and the gray bands in Figs. 4 and 5  below, that represent the $N$-$m$ parameter space. 
The black band represent the self-interacting $\phi^4$ glueball dark matter.
Fig. 4 is based on the $\sigma\to\infty$ GR limit, whereas Fig. 5 takes into account the brane tension bound $\sigma \gtrsim  3.18\times10^6 \;{\rm MeV^4}$. 
 \begin{figure}[H]
\centering\includegraphics[width=8cm]{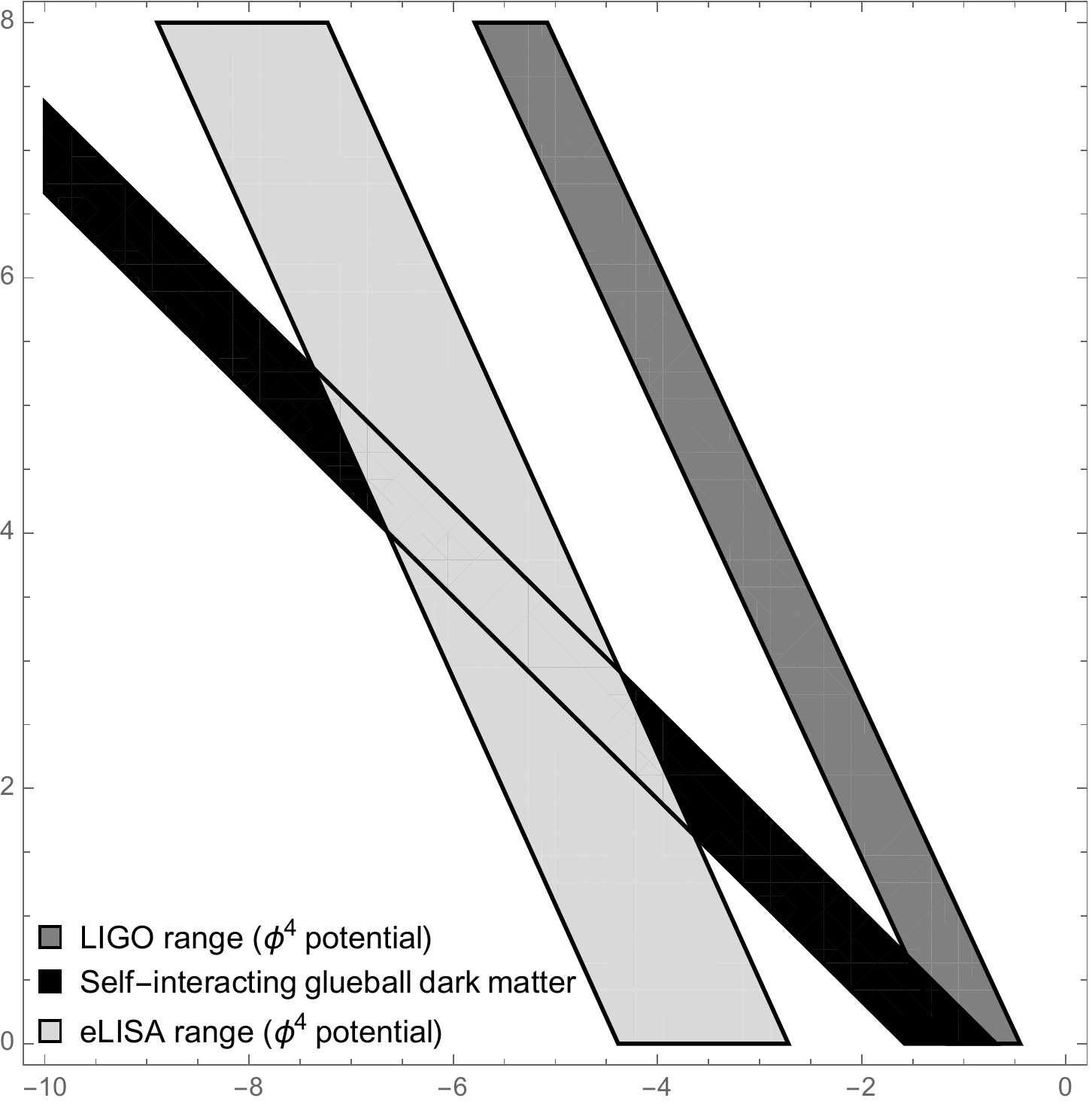}
\caption{The parameter space of $m$ (GeV) versus $N$ for the self-interacting glueball dark matter, using the $\phi^4$ potential, in the $\sigma\to\infty$ GR limit. The gray [light gray] band indicates the  highest frequency of gravitational wave radiation that can be detected by the LIGO [eLISA] experiment. The black band regards the $\phi^4$ self-interacting glueball dark matter.}
\end{figure}
 \begin{figure}[H]
\centering\includegraphics[width=8cm]{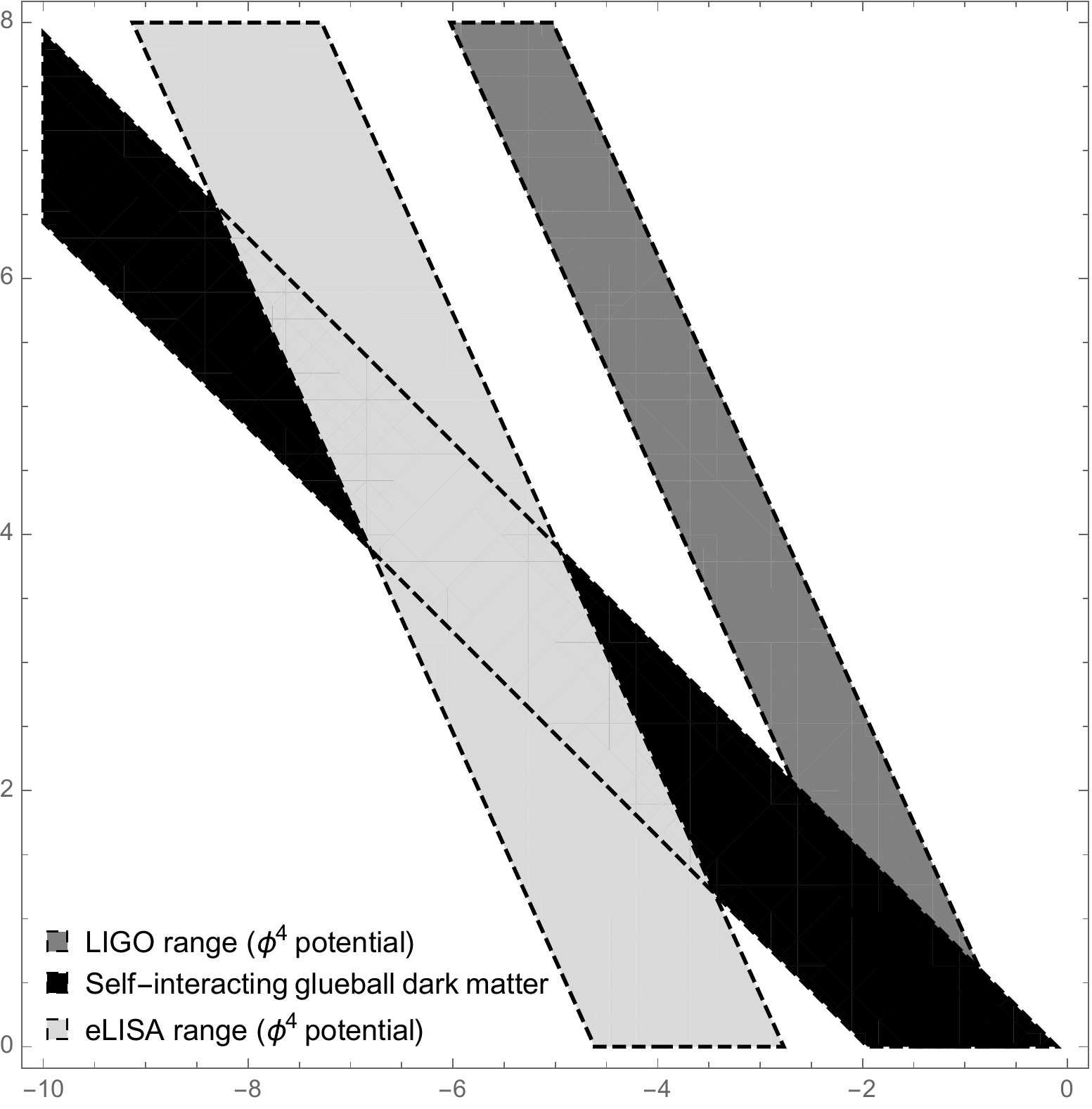}
\caption{The parameter space of $m$ (GeV) versus $N$ for the self-interacting glueball dark matter, using the $\phi^4$ potential, in the current brane tension bound $\sigma \gtrsim  3.18\times10^6 \;{\rm MeV^4}$ \cite{Casadio:2016aum}. The gray [light gray] band indicates the  highest frequency of gravitational wave radiation that can be detected by the LIGO [eLISA] experiment. The black band regards the $\phi^4$ self-interacting glueball dark matter.}
\label{ppppp1}
\end{figure}
The differences between Fig. 4 and Fig. 5, both for the $\phi^4$ glueball potential, reside on the distinction between the GR limit and the MGD setup, respectively. 
Although the spectrum of frequencies 
detected by the LIGO and the eLISA experiments are slightly modified by the 5D Weyl fluid in the MGD setup, when one goes from Fig. 4 to Fig. 5, the self-interacting glueball dark matter (black band) in the $N$-$m$ parameter space  is considerably thickened. 

In the next subsection,  
the glueball potential in the large $N$ regime shall be employed, to derive similar corrections, due to a fluid brane variable tension.

\subsection{Large number of SU($N$) colors}

Regarding the large $N$ limit regime, the scalar glueball potential associated with the SU($N$) Yang-Mills dark sector has power counting $\lambda_{i+2} \sim 1/N^i$, where $i\in\mathbb{N}$ for the cubic and higher order terms in Eq. (\ref{quartic}) \cite{Soni:2016gzf,Juknevich:2009ji,Forestell:2016qhc}. Considering  
all higher order terms, the dark glueball potential (\ref{quartic}) in this regime reads \cite{Soni:2016gzf,Juknevich:2009ji,Forestell:2016qhc},
\begin{equation}\label{largenn}
V(\phi) = \left(\frac{m^2 N}{4\pi}\right)^2  \sum_{i=2}^\infty \frac{1}{j!}\left(\frac{4\pi \phi}{Nm}\right)^j \,,
\end{equation} 
which is the Taylor expansion of the exponential function of the argument $\frac{4\pi \phi}{Nm}$, when its two 
first terms are not taken into account \cite{Soni:2016gzf}. 
Similarly to Eqs.~(\ref{Klein2} --  \ref{einstein2}), the coupled equations can be acquired \cite{Soni:2016yes,Soni:2016gzf}:
\begin{subequations}\begin{eqnarray}
&&\frac{A({\rm x})}{\Omega^2}={ }_2F_1^{-1}\left(0.5; \left\{ 1, 1.5\right\}; 4\pi \upphi^2({\rm x}) \rule{0mm}{4mm}\right)\ , \label{20} \\
&&{\rm M}'({\rm x}) - {\rm x}^2 \left( \frac{\Omega^2}{4A(x)} \upphi^2({\rm x}) + \frac{\mathcal{I}_0}{16\pi^2}  \right) =0\ , \label{21} \\
&&\!\!\!\!\!\frac{\left(\ln A({\rm x})\right)'}{B({\rm x})} - \frac{2 {\rm M}({\rm x})}{{\rm x}^2} - \frac{\Omega^2{\rm x}}{2A({\rm x})}\upphi^2({\rm x}) + 
\frac{{\rm x}\mathcal{I}_0}{8\pi^2}=0 \ ,
\end{eqnarray} 
\end{subequations}
where $\mathcal{I}_0=  {\rm I}_0 \left(\upphi({\rm x}) -1\right)$, and the $\Lambda 
\gg 1$ regime is adopted \cite{Soni:2016yes,Soni:2016gzf}. The  symbol ${ }_2F_1$ is the usual  generalized hypergeometric function and ${\rm I}_0$ denotes the (modified) Bessel function. These two functions
are derived when the time averaging of the potential (\ref{largenn}) is computed \cite{Soni:2016gzf}.

The results obtained from using the large $N$ limit glueball potential in Eq. (\ref{largenn})
 are plot in Figs.~6 - 8. 
 \begin{figure}[H]
\centering\includegraphics[width=7.1cm]{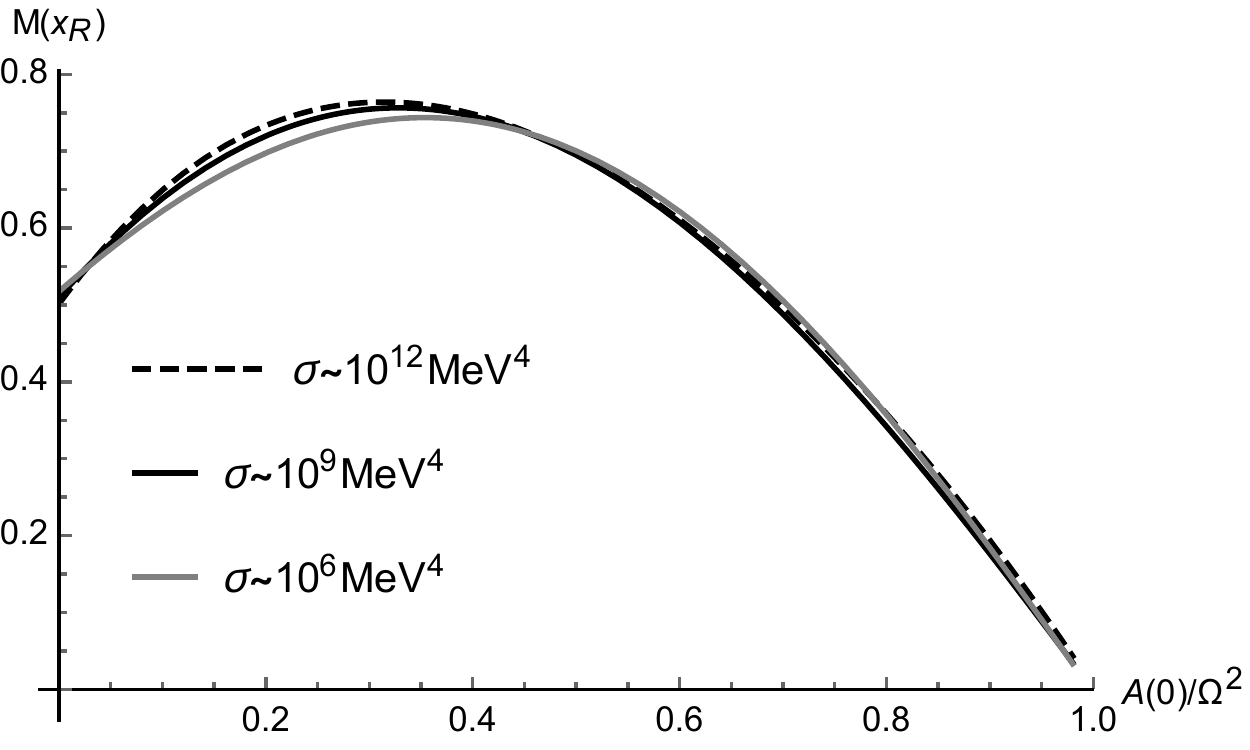}
\caption{Dark SU($N$) MGD star mass ${\rm M}(x_R)$, in the scalar glueball potential (\ref{largenn}), normalized by $\frac{\sqrt{2\lambda}\,M_{\rm pl}^3}{m^2}$, with respect to $\frac{A(0)}{\Omega^2}$, for different values of the fluid brane tension $\sigma = 10^{12}$ MeV$^4$ (dashed line); $\sigma = 10^9$ MeV$^4$ (black line); $\sigma \sim 10^{6}$ MeV$^4$ (gray line).}\centering\includegraphics[width=7.1cm]{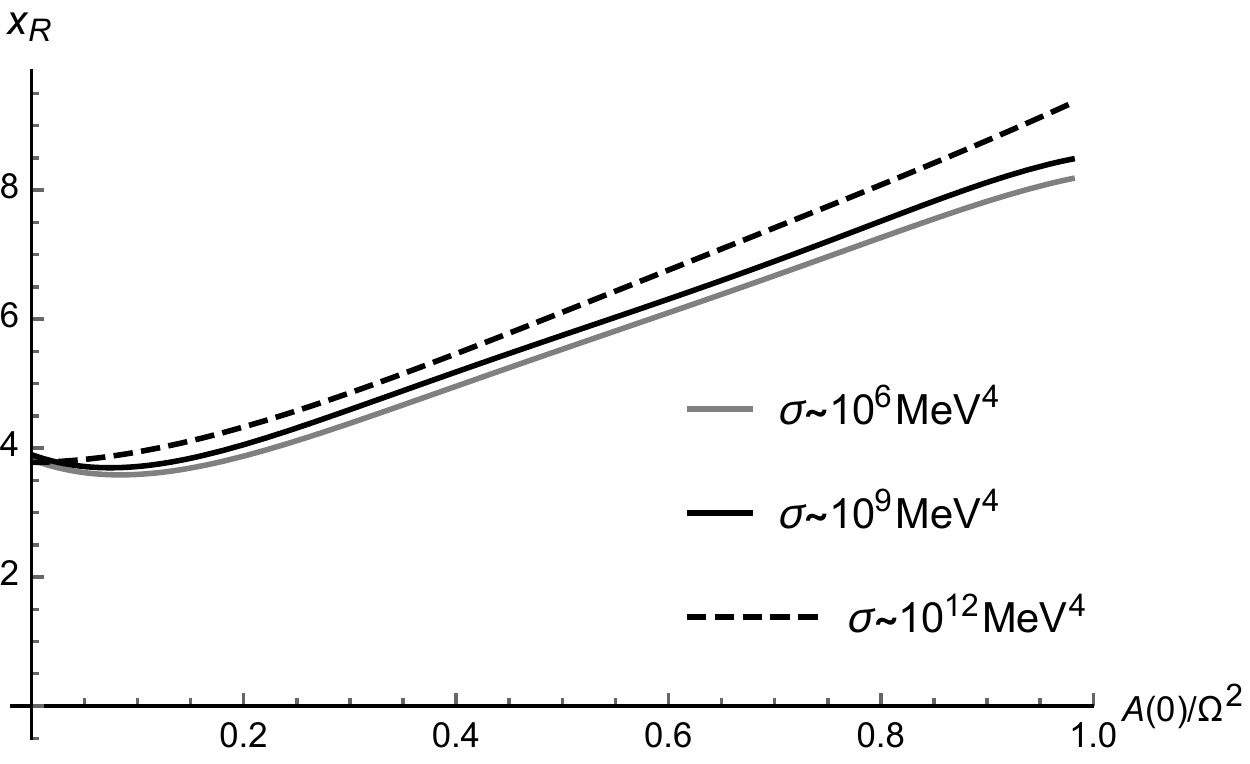}
\caption{Dark SU($N$) MGD star radius $x_R$, in the scalar glueball potential (\ref{largenn}), normalized by $\frac{\sqrt{2\lambda}\,M_{\rm pl}}{m^2}$, with respect to  $\frac{A(0)}{\Omega^2}$, for different values of the fluid brane tension $\sigma = 10^{12}$ MeV$^4$ (dashed line); $\sigma = 10^9$ MeV$^4$ (black line); $\sigma \sim 10^{6}$ MeV$^4$ (gray line).}\centering\includegraphics[width=7.1cm]{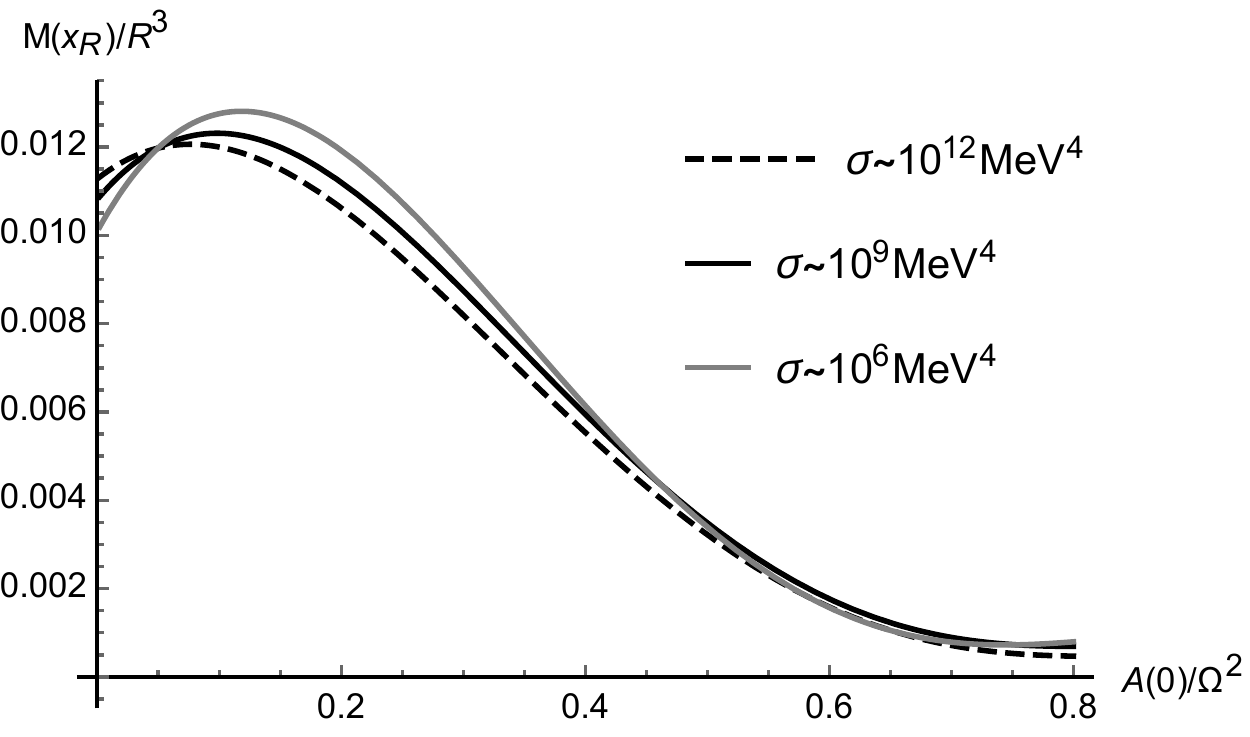}
\caption{Dark SU($N$) MGD star ration $\frac{{\rm M}(x_R)}{x_R^3}$, in the scalar glueball potential (\ref{largenn}), normalized by $\frac{\sqrt{2\lambda}\,M_{\rm pl}}{m^2}$, with respect to $\frac{A(0)}{\Omega^2}$, for different values of the fluid brane tension $\sigma = 10^{12}$ MeV$^4$ (dashed line); $\sigma = 10^9$ MeV$^4$ (black line); $\sigma \sim 10^{6}$ MeV$^4$ (gray line).}\label{ppppp}
\end{figure}
Figs. 6 - 8  take into account a finite brane tension, being smoother
 than their respective $\sigma\to\infty$ counterparts \cite{Soni:2016gzf}.
Comparing Figs. 1, 2 and 3 (the $\phi^4$ dark glueball potential) to, respectively, Figs. 4, 5, and 6 (the large $N$ limit dark glueball potential), one realizes that the dark SU($N$) MGD stars have bigger radii and are more massive that their 
GR limit counterparts. 

 Analogously to the self-interacting $\phi^4$ glueball potential, in Fig. 6   the mass spectrum ($y$-axis) between  the point $\frac{A(0)}{\Omega^2}=0$ up to the critical (maximum) point in the plots cannot be attained, and a dark SU($N$) star accretes and increases its mass up to a maximum, represented in the third column in Table II. The results are presented for different values of the brane tension: 
\begin{center}
\begin{table}[!h]
\begin{tabular}{||c||c||c||c||c||}\hline\hline
\;Brane tension $\sigma$\vspace{0.06cm}\;&\;$
\frac{A(0)}{\Omega^2}$\;&Mass M$(x_R)$& Radius $x_R$&$\frac{{\rm M}(x_R)}{x_R^3}$\;\\\hline\hline
$\infty$ (GR limit)&0.319&0.74&4.7&0.0070\\\hline
$10^{12}$  (MeV$^4$)&0.322&0.74&4.9&0.0073\\\hline
$10^{9}$  (MeV$^4$)&0.327&0.73&4.9&0.0083\\\hline
$10^{6}$  (MeV$^4$)&0.372&0.72&5.0&0.0084\\\hline\hline
\end{tabular}
\caption{highest values of the dark SU($N$) MGD stars radius $\left(\text{normalized by $\frac{\sqrt{2\lambda}}{m^2}\,M_{\rm pl}$}\right)$ and mass $\left(\text{normalized by $\frac{\sqrt{2\lambda}}{m^2}\,M_{\rm pl}^3$}\right)$,  by accretion, for different values of the fluid brane tension, for a the scalar glueball potential (\ref{largenn}).}
\end{table}
\end{center}
 Table II show that the ratio $\frac{{\rm x}_R}{{\rm M}(x_R)}$ is 
always greater than 2 (in normalized units), irrespectively of the brane tension values, whatever the glueball potential is considered. It means that the radius of a dark SU($N$) MGD star is always larger
than a Schwarzschild  black hole event horizon of same mass, a similar result to the one in Ref. \cite{Soni:2016gzf} that considers the $\sigma\to\infty$  GR limit. Hence there is no collapse process of a dark SU($N$) MGD star that originates a black hole. 
 The highest frequency of the gravitational wave radiation is again obtained for the maximum value of the dark SU($N$) MGD star mass. Both the MGD star highest effective radius and mass, respectively,  $R =  \frac{1}{2\sqrt{\pi}} \frac{M_{\rm pl}}{N m^2} x_R$ and $M= \frac{1}{2\sqrt{\pi}}  \frac{M_{\rm pl}^3}{N m^2} {\rm M}(x_R)$ are presented: 
\begin{eqnarray}\label{massa2}
\!R&\!=\!&\! \begin{cases} \frac{3.57}{Nm^2},&\!\!\quad \text{for $\sigma\to\infty$ (GR limit)}\label{mm1}\\
 \frac{3.72}{Nm^2},&\!\!\quad \text{for $\sigma=10^{12}\;{\rm MeV}^4$} \label{mfka2}\\
\frac{3.72}{Nm^2},&\!\!\quad \text{for $\sigma=10^{9}\;{\rm MeV}^4$}\label{mfka3}\\
\frac{3.79}{Nm^2},&\!\!\quad \text{for $\sigma\sim10^{6}\;{\rm MeV}^4$}\label{mfka4}
\end{cases}\\
\!M&\!=\!&\! \begin{cases}\frac{0.36}{Nm^2} M_{\odot},&\!\!\quad \text{for $\sigma\to\infty$ (GR limit)}\label{mm1}\\
\frac{0.36}{Nm^2}   M_{\odot},&\!\!\quad \text{for $\sigma=10^{12}\;{\rm MeV}^4$} \label{mfka2}\\
\frac{0.35}{Nm^2}   M_{\odot},&\!\!\quad \text{for $\sigma=10^{9}\;{\rm MeV}^4$}\label{mfka3}\\
\frac{0.34}{Nm^2}   M_{\odot},&\!\!\quad \text{for $\sigma\sim10^{6}\;{\rm MeV}^4$}\label{mfka4}
\end{cases}.\label{raioo2}\end{eqnarray}

One then gets a highest gravitational wave frequency, that reads
\begin{eqnarray}\label{cccc}
\!\!\!\!\!\!\!\!f_{\rm max} = \frac{m^2N}{\sqrt{\pi} M_{\rm pl}} { {\rm supp\,}^{\!\frac12}\left(\frac{{\rm M}(x_R)}{x_R^3}\right)} 
\simeq  \beta_2(\sigma)\,(50\, {\rm Hz}), 
\end{eqnarray} 
where the function \begin{eqnarray}
\beta_2(\sigma) \approxeq 123.4\,\mathfrak{c} \;(m^2N)\,{\rm GeV}^{-2},
\end{eqnarray} has a tuning factor $\mathfrak{c}$, that is a function of the variable brane tension, according to Eq. (\ref{cccc}) and to the last column of Table II, given by \begin{eqnarray}\label{mfkc}\mathfrak{c}=
\begin{cases}1,&\quad \text{for $\sigma\to\infty$ (GR limit)}\\
1.020,&\quad \text{for $\sigma=10^{12}\;{\rm MeV}^4$}, \\
1.088,&\quad \text{for $\sigma=10^{9}\;{\rm MeV}^4$},\\
1.114,&\quad \text{for $\sigma\sim10^{6}\;{\rm MeV}^4$},
\end{cases}\end{eqnarray}
The parameter $\mathfrak{c}$ indicates 
the corrections to the unit (that corresponds to the $\sigma\to\infty$ GR limit), for different values for the brane tension.

Similarly to what was accomplished in the Subsect. III. A,  dark SU($N$) stars have parameters in the ranges $100$ eV $\lesssim m \lesssim 10\,$KeV and $10^3\lesssim N \lesssim 10^6$, yielding a maximum $10^6M_{\odot}\lesssim{M}\lesssim10^9M_{\odot}$ for the dark SU($N$) MGD star mass, whereas the lowest dark SU($N$) MGD star radius lies  in the range $10^2\lesssim R\lesssim10^5$. 
Hence,  the highest frequency of the gravitational wave by dark SU($N$) MGD star mergers, is ruled by Eqs.~(\ref{cccc}) for the large $N$ limit glueball potential.  The  highest frequency of gravitational wave radiation $f_{\rm max}$, from  dark SU($N$) MGD stellar mergers, can be allocated in the range 30 $\mu${\rm Hz} $\lesssim f_{\rm max}\lesssim 100$ mHz, that can be further detected by the eLISA mission \cite{Seoane:2013qna}. Also, the LIGO experiment  can  probe the spectrum 50 {\rm Hz} $\lesssim f_{\rm max}\lesssim 1$ KHz. Both these spectra are, respectively, represented by the light-gray and the gray bands in Figs. 9 and 10 below,  respectively, in the $\sigma\to\infty$ GR limit and in the brane tension bound $\sigma \gtrsim  3.18\times10^6 \;{\rm MeV^4}$, for the large $N$ glueball potential.

 \begin{figure}[H]
\centering\includegraphics[width=8cm]{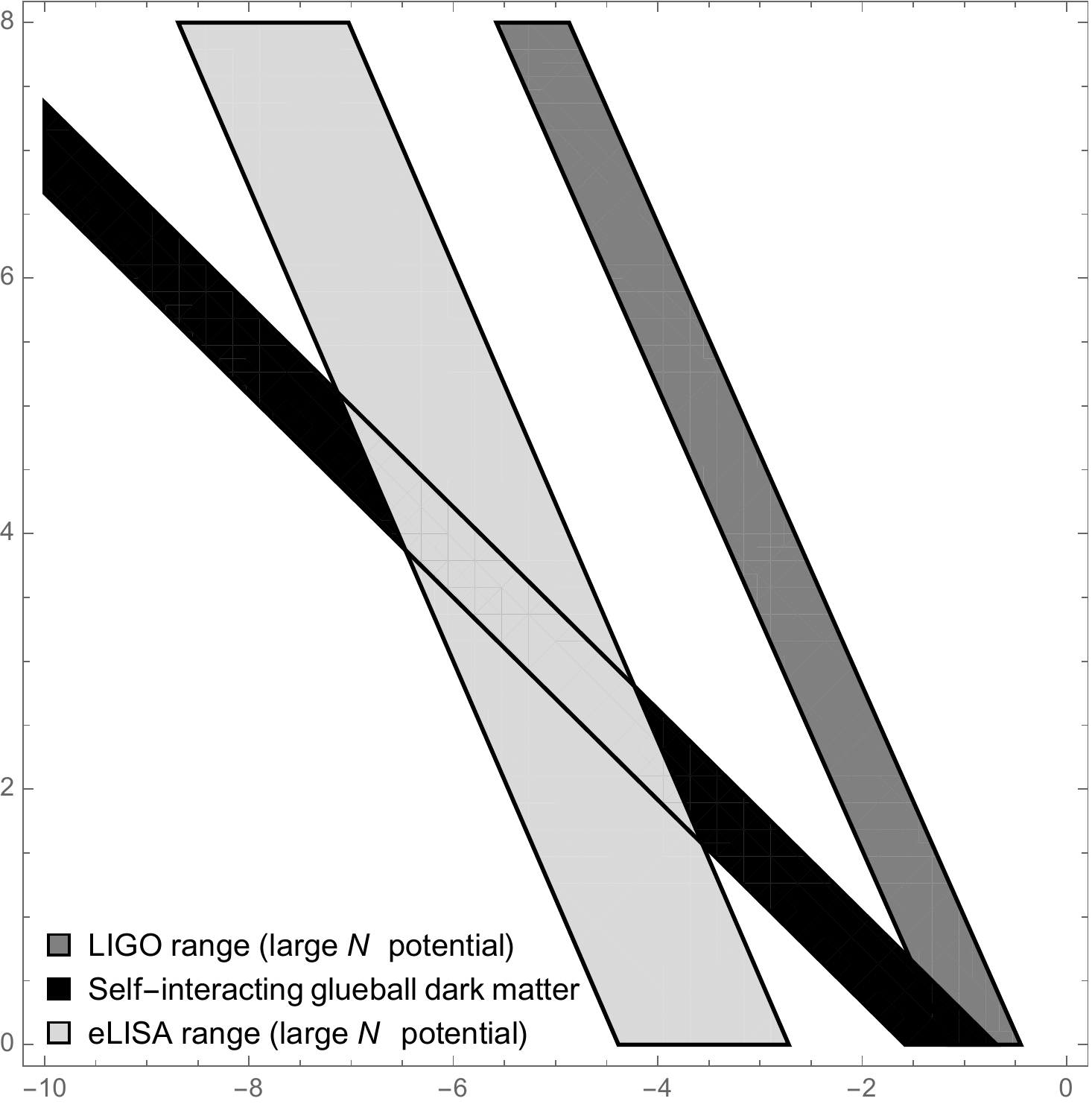}
\caption{The parameter space of $m$ (GeV) versus $N$ for the self-interacting glueball dark matter, using the large-$N$ potential, in the $\sigma\to\infty$ GR limit. The gray [light gray] band indicates the  highest frequency of gravitational wave radiation that can be detected by the LIGO [eLISA] experiment. The black band regards the large $N$ self-interacting glueball dark matter.}
\end{figure}
 \begin{figure}[H]
\centering\includegraphics[width=8cm]{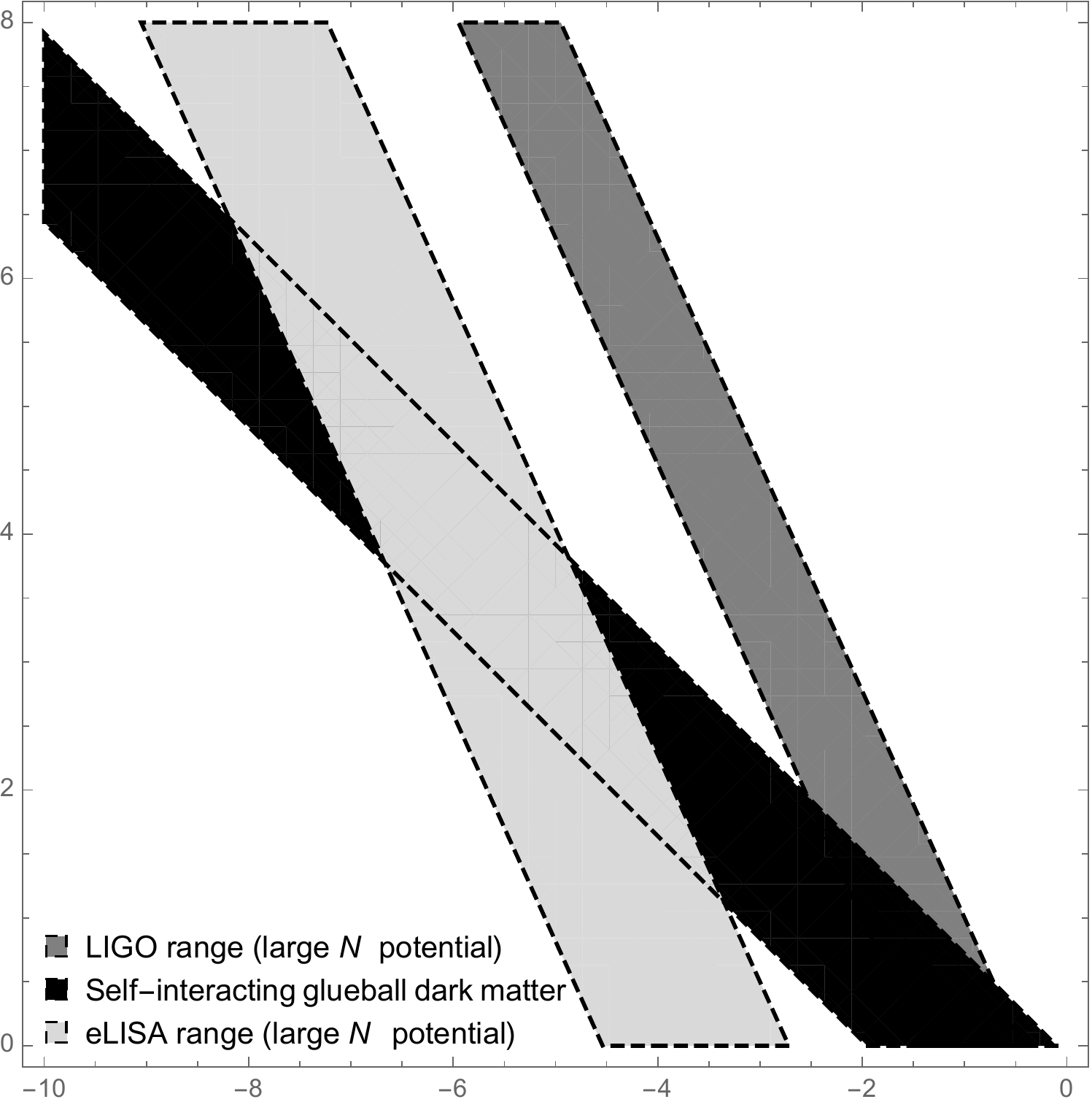}
\caption{The parameter space of $m$ (GeV) versus $N$ for the self-interacting glueball dark matter, using the large-$N$ potential, in the current brane tension bound $\sigma \gtrsim  3.18\times10^6 \;{\rm MeV^4}$ \cite{Casadio:2016aum}. The gray [light gray] band indicates the  highest frequency of gravitational wave radiation that can be detected by the LIGO [eLISA] experiment. The black band regards the large $N$ self-interacting glueball dark matter.}\label{ppppp2}
\end{figure}
Similarly to the case analyzed in Subsect. III. A, the $\sigma\to\infty$ GR limit  has a tiny intersection between the possibility of detection at the LIGO experiment  and the black band that represents the self-interacting glueball dark matter, in Fig. 9. However, the finite brane tension make this black band to be larger in Fig. 10. It represents an increased spectrum to be probed by the LIGO and eLISA experiments, that are themselves also slightly thickened by brane-world effects. It can be realized by respectively comparing the gray and the light-gray bands in Figs. 9 and 10. 

\section{Concluding remarks and outlook}

The pure SU($N$) Yang-Mills glueball dark matter model was proposed to condense into dark SU($N$) MGD stars on fluid branes, representing compact configurations of scalar glueball fields.                The scalar glueball dark matter model may, eventually, decay into Standard Model elementary particles \cite{Juknevich:2009ji}. 
Figs. 5 and 10 shows that a fluid brane-world scenario, wherein the brane tension varies according to Eq. (\ref{tensao}), provides a thickened band that represents the self-interacting glueball dark matter, for respectively both $\phi^4$ and the large $N$ potentials that describe the glueball self-interaction.  Since by self-interaction the glueballs can agglutinate and condense into dark SU($N$) MGD stars, this band has a larger intersection with the LIGO (and the eLISA) experiment to detect gravitational waves. 

Taking into account a finite brane tension  makes the signature of dark SU($N$) star mergers  
 more susceptible to be detected by the LIGO experiment and by the future LISA/eLISA project. In fact, with the  most recent and precise brane tension bound \cite{Casadio:2016aum}, Fig. 5  
 shows a bigger area of intersection,  between the self-interacting $\phi^4$ glueball dark matter and both the experimental windows for detection, than its GR $\sigma\to\infty$ limit depicted in Fig. 4. Analogously, for the large $N$ potential, 
 Fig. 10 also presents a larger range of intersection,  between the large $N$ self-interacting potential glueball dark matter and both the LIGO and the eLISA detectable spectra.
 To summarize, Figs. 4 and 9 represent the GR $\sigma\to\infty$ limit, respectively both $\phi^4$ and the large $N$ potentials that describe the glueball self-interaction, representing a more improbable scenario to detect gravitational waves. 
Hence, dark SU($N$) MGD stars,  that have mass and radius corrected by the the bulk 5D Weyl fluid, should be better detectable by the current LIGO experiment \cite{Abbott:2007kv} and by the eLISA project \cite{Seoane:2013qna}. 
 
Finally, the search for TeV-scale gravity signatures in the ATLAS detector at $s=13$ TeV is being currently approached \cite{Aaboud:2016ewt}. 
Black holes produced with a mass above the formation threshold evaporate in Higgs particles, leptons, particle jets and photons and are currently searched at the LHC. Moreover, signatures of TeV extra-dimensional models  are encoded in the partners of the Z and W bosons that might be permitted to access the 5D bulk. These partners can be manifested as resonances in dilepton spectra and are more intricate to detect at the LHC. Kaluza-Klein partners of the Standard Model particles have not been observed still, pushing the lower mass limits beyond 13 TeV.
To summarize, collider data indicate string theory magnitude extra dimensions, whose searches   continue at the LHC and confirmation could happen in the next generation of colliders \cite{Deutschmann:2017bth}.  The search for evidences of extra dimensions can be, then, dislocated also to observational aspects, as accomplished in Sect III, by investigating models whose signatures increase the spectrum of the highest frequency of gravitational wave radiation  by the LIGO and eLISA experiments.\subsection*{Acknowledgements}

RdR~is grateful to CNPq (Grant No. 303293/2015-2),
and to FAPESP (Grant No.~2015/10270-0), for partial financial support. 

\bibliography{bib_DSS}

\end{document}